\documentclass[10pt]{iopart}

\usepackage{graphicx}	
\usepackage[utf8]{inputenc} 	
\usepackage[T1]{fontenc} 	

\newcommand{\lint}{L_{c}}
\newcommand{\tfix}{t_{\textrm{\tiny fix}}}
\newcommand{\tmut}{t_{\textrm{\tiny mut}}}

\newcommand{\etwait}{\langle t_{{k}}\rangle}

\begin{document}
\title[Spatial structure increases the waiting time for cancer]
{Spatial structure increases the waiting time for cancer}

\author{Erik A. Martens$^1$, Rumen Kostadinov$^2$, Carlo C. Maley$^3$ and Oskar Hallatschek$^1$}
\address{$^1$ Max Planck Research Group for Biophysics and Evolutionary Dynamics, MPI for Dynamics and Self-Organization, G\"{o}ttingen, Germany}
\address{$^2$ School of Medicine, University of Pennsylvania, Philadelphia, USA}
\address{$^3$ Center for Evolution and Cancer, Helen Diller Family Comprehensive Cancer Center, and Department of Surgery, University of California, San Francisco, USA}
\ead{erik.martens@ds.mpg.de, rkostadi@gmail.com, carlo.maley@ucsf.edu and oskar.hallatschek@ds.mpg.de}

\begin{abstract}
Cancer results from a sequence of genetic and epigenetic changes which lead to a variety of abnormal phenotypes including increased proliferation and survival of somatic cells, and thus, to a selective advantage of pre-cancerous cells.
The notion of cancer progression as an evolutionary process has been experiencing increasing interest in recent years. 
Many efforts have been made to better understand and predict the progression to cancer using mathematical models; these mostly consider the evolution of a well-mixed cell population, even though pre-cancerous cells often evolve in highly structured epithelial tissues. In this study, we propose a novel model of cancer progression that considers a spatially structured cell population where clones expand via adaptive waves.
This model is used to assess two different paradigms of asexual evolution that have been suggested to delineate the process of cancer progression. The standard scenario of periodic selection assumes that driver mutations are accumulated strictly sequentially over time.  However, when the mutation supply is sufficiently high, clones may arise simultaneously on distinct genetic backgrounds, and clonal adaptation waves interfere with each other. We find that in the presence of clonal interference, spatial structure increases the waiting time for cancer, leads to a patchwork structure of non-uniformly sized clones, decreases the survival probability of virtually neutral (passenger) mutations,
and that genetic distance begins to increase over a characteristic length scale $L_c$. 
These characteristic features of clonal interference may help to predict the onset of cancers with pronounced spatial structure and to interpret spatially-sampled genetic data obtained from biopsies.
Our estimates suggest that clonal interference likely occurs in the progression of colon cancer, and possibly other cancers where spatial structure matters. 
\end{abstract}


\submitto{\NJP}
\maketitle
\section{Introduction}
Progression to cancer is a process of somatic evolution within the body. The idea that cancer can be understood as an evolutionary process was already formulated by Nowell in 1976~\cite{Nowell1976}; still, relatively little attention has been directed on applying principles from population genetics to understand and control neoplastic progression until recently~\cite{Tsao2000,Gonzalez-Garcia2002,Maley2006,Beerenwinkel2007,Bozic2010,Michor2004}. 

The progression to cancer may be seen as a multi-stage process~\cite{Frank2007book}, where 
the various stages are characterized by different phenotypic outcomes conferred by somatic evolution. 
Many cancers are typically initiated with a benign transformation of tissue that later progresses to invasion (malignancy).
For instance, in colon cancer, the observation of low and then high-grade dysplasia (early and late adenoma) is followed by the formation of a carcinoma~\cite{Jones2008}. Similarly, Barrett's esophagus is a condition where normal squamous esophageal tissue has transformed into cells of intestinal type (metaplasia), a change associated with chronic gastric reflux; later, the tissue may proceed to low and then high-grade dysplasia, before an actual carcinoma develops~\cite{Reid2010}. 
The waiting time that passes from an initially benign tissue transformation until the formation of malignant neoplasm (conferring cancerous growth) can last for a few up to 20 years~\cite{Jones2008}. 
Many neoplasms, however, never progress to malignancy during the lifetime of a patient.\footnote{For instance, ulcerative colitis is an inflammatory bowel disease, in which only 10\% of the patients eventually develop cancer~\cite{Salk2009}.} The determination of which patients are at risk remains an open problem. Improved models are required to explain the progression to cancer, assisting us in interpreting genotyping data from biopsies, and in developing improved diagnostic tools and interventions.

Here, we are concerned with the genotypic evolution of cancer, rather than with phenotypic outcomes. Tumorigenesis occurs by accumulation of mutations in oncogenes, tumor suppressor genes, and genes causing genetic instability in neoplasms~\cite{Vogelstein2004}. Neoplasms are abnormal tissue masses with increased proliferation of cells and may be benign or malignant. A sequence of mutations (or ''hits``, which may include chromosomal alterations and epigenetic changes, not just DNA sequence mutations) may lead to the typical hallmarks of cancer\footnote{Hallmarks of cancer include self-sufficiency in growth signals, insensitivity to anti-growth signals, tissue invasion and metastasis, limitless reproductive potential, sustained angiogenesis, evading apoptosis and possibly others~\cite{Hanahan2000,Hanahan2011}.}.
From an evolutionary perspective, a neoplasm can be viewed as a large, genetically  heterogeneous population of cells, which acquire heritable genetic and epigenetic alterations, influencing their survival and reproduction. The fitness of a neoplastic cell is shaped by its interactions with cells and other environmental factors. Indeed, many of the (phenotypic) alterations proposed as hallmarks of cancer provide a selective advantage to the neoplastic cells~\cite{Hanahan2000}. 
Genetic alterations that are advantageous for a neoplastic cell are normally deleterious to the host and ultimately cause death to both host and neoplasm. 
Because somatic abnormalities have differing, heritable effects on the fitness of neoplastic cells, mutant clones may expand or decrease in size by the principles of Darwinian evolution, e.g.~via natural selection and genetic drift (number fluctuations). 
Clonal evolution generally selects for increased proliferation and survival, and may lead to phenotypes such as invasion, metastasis and therapeutic resistance.

Many  mutations seen in neoplasms seem to be selectively neutral, or even deleterious, while other, so-called driver mutations, confer a selective advantage~\cite{Wood2007,Merlo2006,Kimura1968,Jones2008a,Sjoeblom2006,Parsons2008,Greenman2007}.
Because new mutations only survive genetic drift with a probability proportional to their selective advantage, deleterious (and neutral) mutations likely go extinct. 
However, large numbers of neutral mutations (so-called passenger mutations~\cite{Smith1974}) could avoid extinction by ''hitchhiking`` on the genetic background of selectively advantageous mutations~\cite{Maley2004}. 

While some studies have proposed that only a small number of hits, between 2 and 5, may be required to develop cancer~\cite{Armitage1954,Knudson1971}, some recent studies on breast and colon cancer have suggested that the number of hits may be even larger with up to 20 hits~\cite{Sjoeblom2006,Beerenwinkel2007}. A plausible definition for the waiting time for cancer is the time from the first presence of a neoplasm, until $k$ hits have been accumulated in at least one cell of the neoplasm~\cite{Beerenwinkel2007}.

\begin{figure}[htp]
\centering
    \includegraphics[width=0.45\textwidth]{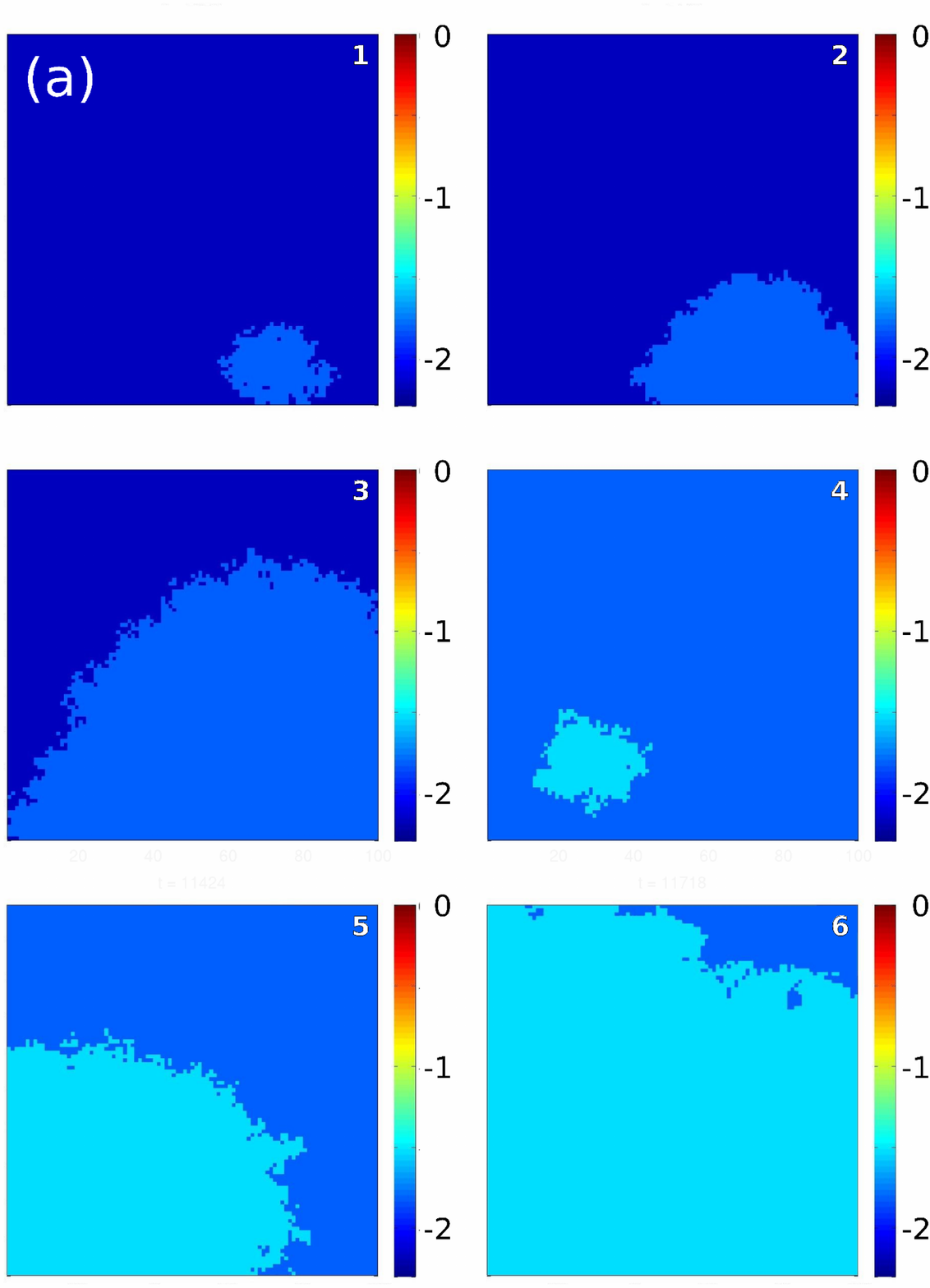} 
    \hspace{.5cm}
    \includegraphics[width=0.45\textwidth]{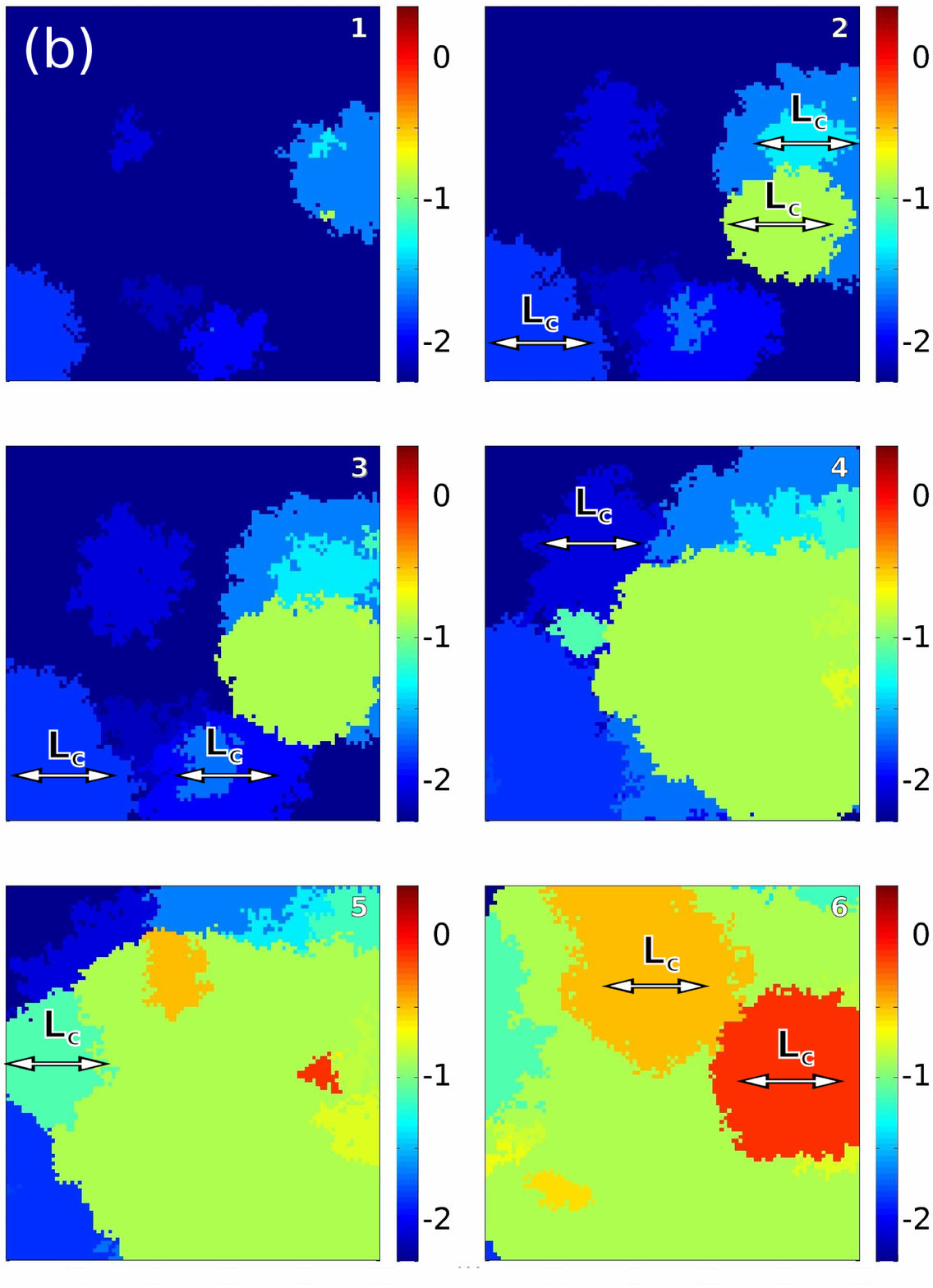} 
    \caption{Two important dynamical regimes may dominate clonal expansions: 
    {\bf Periodic selection (a):} Mutations occur so rarely that adaptive (clonal) waves sweep through the habitat one-by-one.
    {\bf Clonal interference (b):} Frequent mutations lead to simultaneous competition of clones, which collide at a characteristic ''interference'' length $\lint$. In the presence of such clonal interference, only few clones may reach fixation, leading to the loss of clones carrying a selective advantage, thus slowing down evolution~\cite{Desai2007a,Martens2011}.
    Simulations were carried out on a hexagonal lattice with side length $L=100$ and absorbing boundary conditions with a selective advantage of $s_0=0.25$ and a mutation rate of $\mu = 10^{-7}$ in (a), and $\mu = 10^{-5}$ in (b). The interference length is determined via Eq.~(\ref{eq:Lc}) and is $L_c\sim29$ for periodic selection and $L_c\sim 135$ for clonal interference, respectively. Selective advantages are drawn from an exponential distribution. Small numbers denotes the time sequence. Colors define different levels of (logarithmic) fitness. 
    }
\label{fig:PSCI}

\end{figure}

Two contrasting paradigms have been proposed for how driver mutations are accumulated over time:
Periodic selection (Fig.~\ref{fig:PSCI}(a)) occurs when the waiting time for a successful mutation is much longer than its fixation time, i.e.~the time it takes a clone to spread throughout the entire neoplasm: in this case, clones expand strictly sequentially. 
Vice versa, when the typical fixation time is much larger than the waiting time for the next successful mutation, multiple clones, arising on different genetic backgrounds, may compete with one another to reach fixation (Fig.~\ref{fig:PSCI}(b)). In this case, clones collide at a characteristic ``interference'' length $L_c$~\cite{Martens2011}, which we discuss in detail in the Results section. 
Due to this clonal interference, only a small number of selectively advantageous mutations reach fixation, while most are lost, thus leading to a reduced speed of evolution when compared to periodic selection. 
The model of periodic selection has long shaped the field of population genetics, partly because beneficial mutations were thought to be exceedingly rare.
However, in recent years, clonal interference has been found to be of great importance 
in experimental studies~\cite{Desai2007b} on microbial evolution. While microbes certainly are quite different from neoplastic cells, both paradigms of periodic selection and clonal interference have been suggested to be of significance in describing the progression and dynamics of cancer~\cite{Maley2004,Leedham2008,Graham2010}.

\begin{figure}[htp]
\centering
    \includegraphics[width=0.9\textwidth]{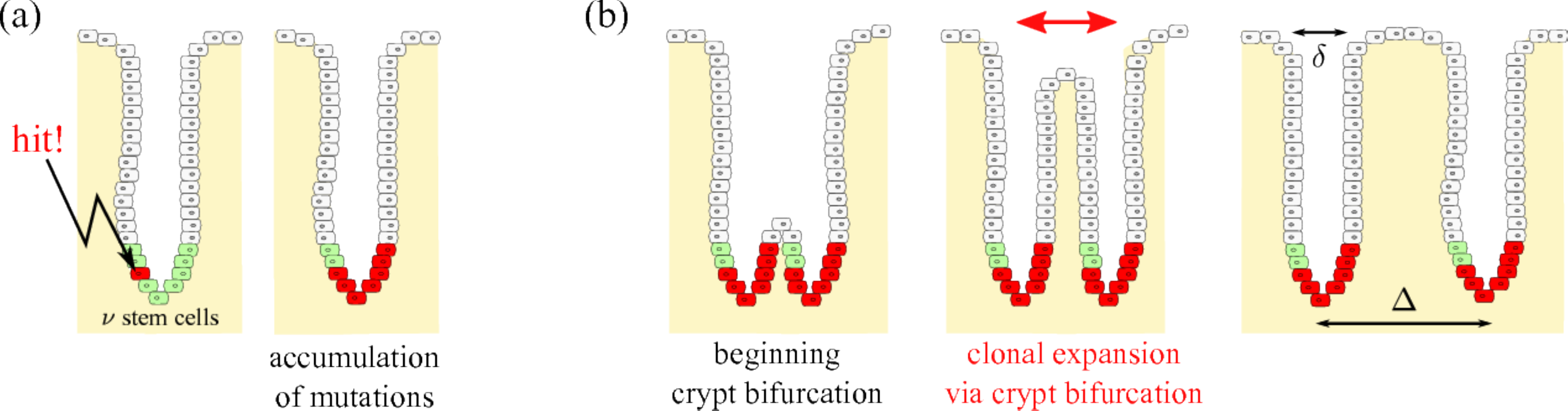}
    \caption{Clonal expansion is possible via crypt bifurcation, and has been observed in the human colon~\cite{Salk2009,Greaves2006} and may play an important role in Barrett's esophagus~\cite{Reid2010}. Mutations can occur anywhere within the epithelium but persist only in stem cells. The number $\nu$ of stem cells in a crypt is very small. A mutation can spread in the stem cells until most stem cells within the crypt carry that mutation (a). At some point, crypts go through a bifurcation process (b). Such processes continue throughout the life of the host. If the mutation is pro-oncogenic, it may lead to a higher proliferation rate and to the spreading of potentially cancerous mutations over several crypts in the neoplasm. Crypts have an orifice width $\delta$ and are relatively densely packed, with an average spacing $\Delta$.
    }
\label{fig:cryptbifurcation}   
\end{figure}

The dispersal of an allele in cancer can generally happen in three ways: i) cells move between partially isolated sub-populations of proliferative units, ii) locally invade neighboring tissue, or iii) emigrate as metastatic cells from the primary tumor. Invasion and metastasis do not occur until the late stage of cancer progression, where cancerous, abnormal growth is initiated; since we study the progression until cancer initiation, we focus here on the movement of cells between proliferative units (i).
Intestinal epithelium, as well as the epithelium of Barrett's esophagus, is organized in proliferative units called crypts, see Fig.~\ref{fig:cryptbifurcation}(a). 
Intestinal crypts are thought to contain only around 8 to 20 stem cells, thus yielding quite a small effective population~\cite{Calabrese2010}. Stem cells are long-lived and reside close to the bottom of the crypt where they keep renewing the crypt. Thus, crypts sub-divide the epithelium into isolated sub-populations. 
Over time, stem cells acquire mutations which at some point may dominate the stem cell population within a crypt.

In certain pre-cancerous conditions, mutant clones can expand over many crypts~\cite{Salk2009,Maley2004,Greaves2006,Franklin1997}. The mechanisms underlying such clonal expansion are not well understood~\cite{Merlo2006}. Hypotheses include crypt bifurcation (Fig.~\ref{fig:cryptbifurcation}(b)), wounding with epithelial restitution, dispersal through the basement membrane and stroma into the base of neighboring crypts, and the more speculative dispersal over the surface of the epithelium ('local metastasis')~\cite{Merlo2006}. 
However, recent research~\cite{Salk2009,Greaves2006} provide {\it in vivo} evidence for the development of clonal patches (clusters) in the human colon, and suggests that crypt bifurcation indeed is the mechanism explaining clonal expansion in pre-malignant epithelial tissue. In particular, it has been shown that in humans, mutated crypts are clustered together throughout the colon, and that clonal patches increase in size as patients age~\cite{Greaves2006}.
For the purpose of this study, we consider crypt bifurcation as the mechanism behind clonal expansions.


Evolution in spatially structured populations is significantly different from evolution in non-structured populations~\cite{Martens2011}.
A crucial feature of adaptation in spatially structured populations is that advantageous mutations spread in the form of waves, which were first described by R.~A.~Fisher~\cite{Fisher1937,Kolmogorov1937}. 
Because these adaptation waves spread with a constant speed, the width of clones grows linearly with time~\cite{Chao2008};
in contrast, this is much slower than the characteristic exponential growth of Malthusian sweeps in well-mixed environments. Since sweeps are much slower in spatial environments, growing clones are much more likely to interfere than in well-mixed populations. In particular, as fixation times are larger in spatially extended populations, the regime of clonal interference is inflated in structured populations compared to well-mixed populations. As a consequence, the accumulation rate of beneficial mutations (adaptation speed)
is drastically slowed down~\cite{Martens2011}. In particular, it has been shown that, while the adaptation speed in non-structured populations depends logarithmically on population size and mutation rate, the adaptation speed of spatially structured populations saturates for large population sizes, thus representing a strict speed limit of evolution~\cite{Martens2011}. 
Because spatial structure may slow down evolution, we expect that the waiting time for cancer, i.e.~the time until $k$ hits have occurred in a neoplastic cell, is also larger in spatially structured neoplastic tissues. 

Various model approaches have been used to estimate the waiting time for cancer~\cite{Beerenwinkel2007,Bozic2010,Calabrese2010,Foo2011,Meza2008}. 
To our knowledge, so far no model has explicitly accounted for the spatial structure of neoplasms, where adaptive waves describe the expansion of interfering clones.
We thus extend previous work~\cite{Beerenwinkel2007} by using a novel model of spatial evolution~\cite{Martens2011} to study the waiting time for cancer.
Using parameter values found in literature, we estimate that clonal interference is an adequate description of clonal expansion in cancers with spatial structure. Furthermore, we discuss some characteristic features of periodic selection and clonal interference which may help to discern the two competing paradigms~\cite{Graham2010} in experimental data.


\section{Model}

In the following, we introduce a model which allows for an efficient simulation of many clonal expansions simultaneously in progress~\cite{Martens2011,Gordo2006}. The neoplastic tissue is modeled by a hexagonal lattice of constant size $N=L\times L$ with absorbing  boundaries. Each lattice site records the identity of the locally dominating genotype in a crypt.

Each simulation is initiated with a crypt population devoid of mutations.
New mutations carrying a selective advantage appear at a rate $\mu$ per crypt and crypt cycle ($\mu$ thus represents the combined mutation rate of all stem cells harbored in the crypt). When a mutation becomes established at position $i$, the fitness $W_i$ of the $i$-th clone is updated according to $W_i(t+1)= W_i(t)\cdot (1+s)$. Epistasis is absent. The neoplasm is assumed to be homogeneous implying selective pressures throughout the crypt population. The selective fitness advantage $s$ of a mutation is drawn from an exponential distribution with mean value $s_0$.  
New mutations survive genetic drift (stochastic number fluctuations) with a probability $2s$ for small selective advantages $s$. This relation has been shown to hold for both non-structured~\cite{Haldane1927,Kimura1962} and spatially structured populations ~\cite{Martens2011,Maruyama1974,Gordo2006,Goncalves2007} under certain simplifying assumptions.

Natural selection is implemented as follows. A crypt at site $i$ in the offspring generation is replaced by one of its neighboring crypts $j$ from the parent generation with a probability proportional to its fitness $W_j$.
This has the consequence that, at the front between two clones with differing genotypes and fitness difference $\Delta W$, the fitter clone moves into the territory of the other with a velocity $c=c(\Delta W)$.
The replacement rule we use here generates adaptive waves traveling at an average velocity of  $c = (1/4) \sqrt{\Delta W}$~\cite{Martens2011}. This relation is consistent with a model due to R.~Fisher~\cite{Fisher1937}; he found that in a spatially extended habitat where individuals migrate at a rate $m$ between sub-populations, advantageous alleles spread with a velocity of $c=2\sqrt{m \Delta W}$~\cite{Fisher1937,Kolmogorov1937}.

To compare waiting times between spatially structured and non-structured populations, we have adjusted the algorithm to allow for simulations of an effectively well-mixed crypt population: 
instead of selecting the offspring (crypt) from the nearest neighbors of a crypt at site $i$, all neighbors are now replaced with crypts chosen randomly from any crypt neoplasm. Selection is then carried out as described above, i.e.~offspring is selected proportional to the fitness advantages of the randomly chosen competing crypts.

In our simulations, we vary the beneficial mutation rate $\mu$ per site, the mean selective fitness advantage $s_0$, and the side length $L$ of the crypt population. 
Several variables are measured during or at the end of the simulation. 
All simulations are terminated once $k$ mutations have occurred in at least one cell; the simulation time thus represents the waiting time $t_k$ for $k$ hits. We also measure the clonal patch size which we define as the number of crypts with the same mutation. The speed of adaptation (rate of fitness increase) is defined as $V=\langle{\log{{W}(t+1)}}\rangle- \langle{\log{W(t)}}\rangle$, where $\langle\log{W}\rangle$ represents the population average of the logarithmic fitness. We also quantify the spatial genetic (Hamming) distance as a function of spatial distance, as we explain further below.

\section{Results}
\subsection{Distribution of the waiting time to cancer}
\begin{figure}[htp!]
\centering
 \includegraphics[width=0.9\textwidth]{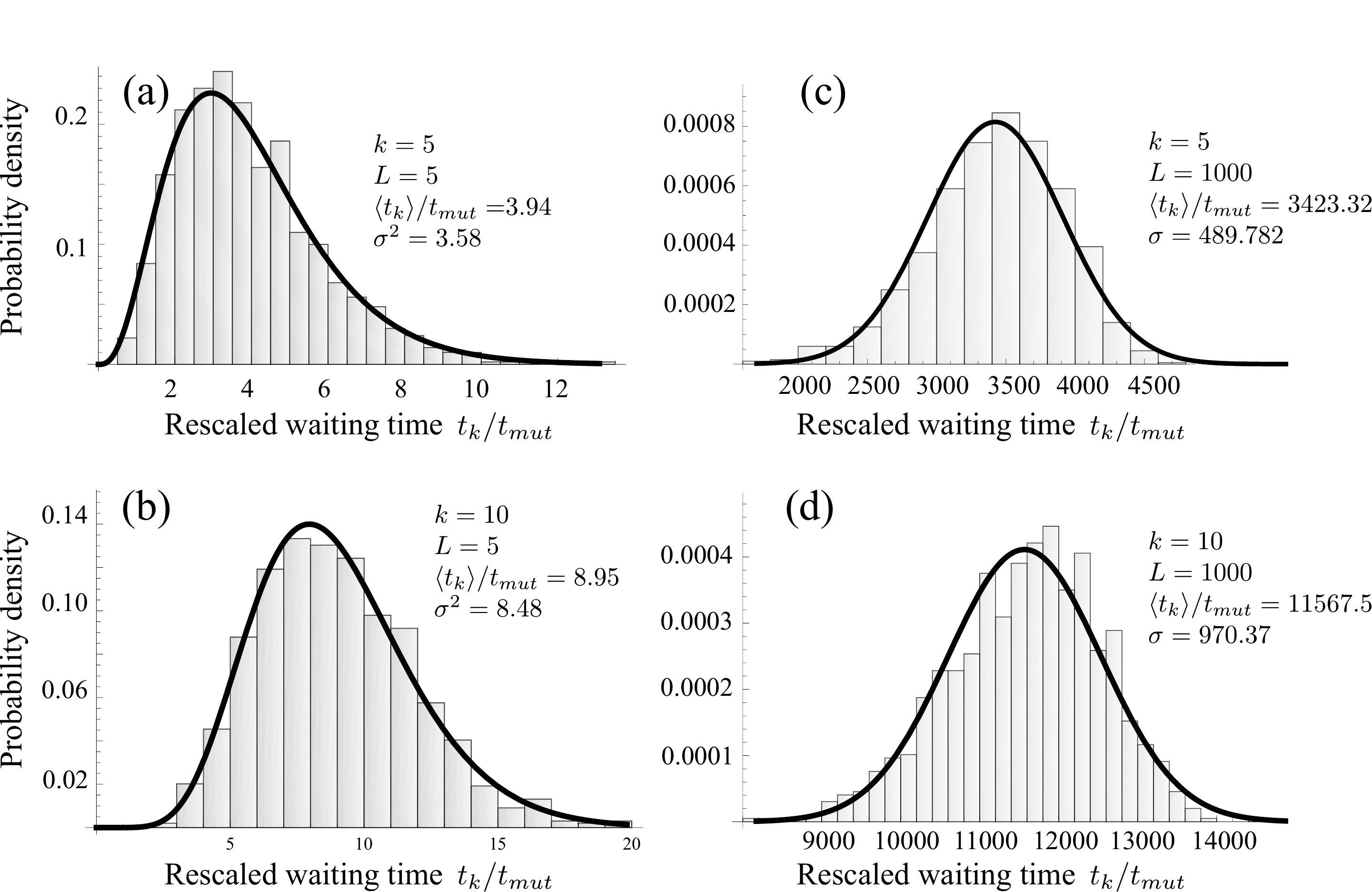}
  \caption{The (rescaled) waiting times for $k$ hits, $t_k/\tmut$, (histogram bars) for small and large neoplasm sizes $L$. When the neoplasm size is small (a,b), clones reach fixation one-by-one. In this case of periodic selection, waiting times are distributed according to Eq.~(\ref{eq:distro1}) (solid curve). 
  Larger neoplasms increase the mutation supply rate, $\mu L^2$, and shorten the typical waiting time for successful mutations to occur, $\tmut=(2s_0\mu L^2)^{-1}$. 
  When the neoplasm size is sufficiently large (c,d), mutations arise frequently and their clones interfere at a characteristic interference length $L_c$, given by Eq.~(\ref{eq:Lc}). However, in this case of clonal interference (c,d), waiting times are normally distributed (solid curve; see text). 
  Data represents 1000 realizations with exponentially distributed fitness effects with mean $s_0=0.05$ and mutation rate $\mu=10^{-4}$ per crypt and cycle, yielding an interference length  of $L_c\sim17$, in agreement with our estimates for colon cancer (Table~\ref{tab:param}).}
\label{fig:twaithisto} 
\end{figure}

Fig.~\ref{fig:twaithisto} shows distributions of waiting times for $k=5$ and $k=10$ hits, for neoplasms of sizes $L=5$ and $L=1000$ and a mutation rate $\mu=10^{-4}$ per crypt and crypt cycle.
When the mutation supply rate for the neoplasm, $\mu L^2$, is sufficiently small, mutations occur so rarely that clones expand strictly sequentially and fixate one-by-one. 
Since new mutations only survive genetic drift with probability $2s$~\cite{Martens2011,Kimura1962,Maruyama1974}, the typical waiting time for one mutation to occur is given by $\tmut=(2s_0\mu L^2)^{-1}$. 
The expected waiting time for $k$ hits is then $\etwait=k\cdot\tmut$, or 
\begin{eqnarray}\label{eq:speedPS}
\etwait &=& k/v,\
\end{eqnarray}
where we introduce the mutation accumulation rate $v$, i.e.~the rate at which hits are acquired; for periodic selection, it is given by $v=1/\tmut$.

In the present case of periodic selection, the accumulation of hits may be understood in terms of $k$ Poisson processes, as explained in~\ref{appA}. The resulting distribution of waiting times for $k$ hits is
\begin{eqnarray}\label{eq:distro1}
 p_k(t) &=&\frac{t^{k-1}e^{-t}}{(k-1)!}.\
\end{eqnarray}
The rescaled waiting times $t_k/\tmut$ follow this distribution (solid curve) for small neoplasms with $L=5$ very closely (Fig.~\ref{fig:twaithisto} (a,b)).

Quite a different picture is obtained for large neoplasms of size $L=1000$, see Fig.~\ref{fig:twaithisto} (c) and (d). Waiting times follow a different distribution than for small neoplasms, and more importantly, the rescaled expected waiting time $\etwait/\tmut$ is several order of magnitudes larger than for the small neoplasm. How can we explain this qualitative and quantitative difference?

The typical time it takes a clone to reach fixation without clonal interference is $\tfix=L/c_0$, where $c_0$ is the speed of a clonal wave with selection coefficient $s_0$.
When this fixation time is much shorter than the waiting time for the next successful mutation to arise, $\tfix\ll\tmut$, adaptive waves expand strictly sequentially (periodic selection). In the opposite case, $\tfix\gg\tmut$, the fixation time is too long for a clone to reach fixation before other mutations occur on a different genetic background, and clones interfere (clonal interference). The cross-over between these two regimes is given by the condition
\begin{eqnarray}\label{eq:crossover}
 (2s_0\mu L_c^2)^{-1} = \tmut \sim \tfix = L_c/c_0.\
\end{eqnarray}
The transition from periodic selection to clonal interference may be achieved by varying neoplasm sizes, as is done in Fig.~\ref{fig:twaithisto}: for sufficiently large neoplasms ($L=1000$), the mutation supply rate becomes large, too, until the condition for clonal interference is fulfilled, i.e.~$\tmut\ll \tfix$. This suggests that there is a critical neoplasm size $L$ at which the cross-over occurs: solving Eq.~(\ref{eq:crossover}) yields the characteristic interference length scale~\cite{Martens2011},
\begin{eqnarray}\label{eq:Lc}
 L_c &=& \left( \frac{c_0}{2s_0\mu}\right)^{1/3}.\
\end{eqnarray}
The length $L_c$ has an intuitive interpretation as the typical length at which clonal waves typically collide: when the neoplasm size $L$ is smaller than $L_c$, waves do not collide (periodic selection); vice versa, we have clonal interference when $L\gg L_c$. The ratio $L/L_c$ compounds the model parameters and represents a measure for the strength of clonal interference (see  Fig.~\ref{fig:twait}).
In the simulations shown in Fig.~\ref{fig:twaithisto}, the characteristic interference length is $L_c \approx 17$; thus, we observe periodic selection when $L=5$ and clonal interference when $L=1000$. 
As we discuss below, the relevant bell-like shape of the (Gaussian) distribution of waiting times for large neoplasms ($L=1000$) is related to the interference of clones.

\begin{figure}[htp!]
\centering
 \includegraphics[width=0.75\textwidth]{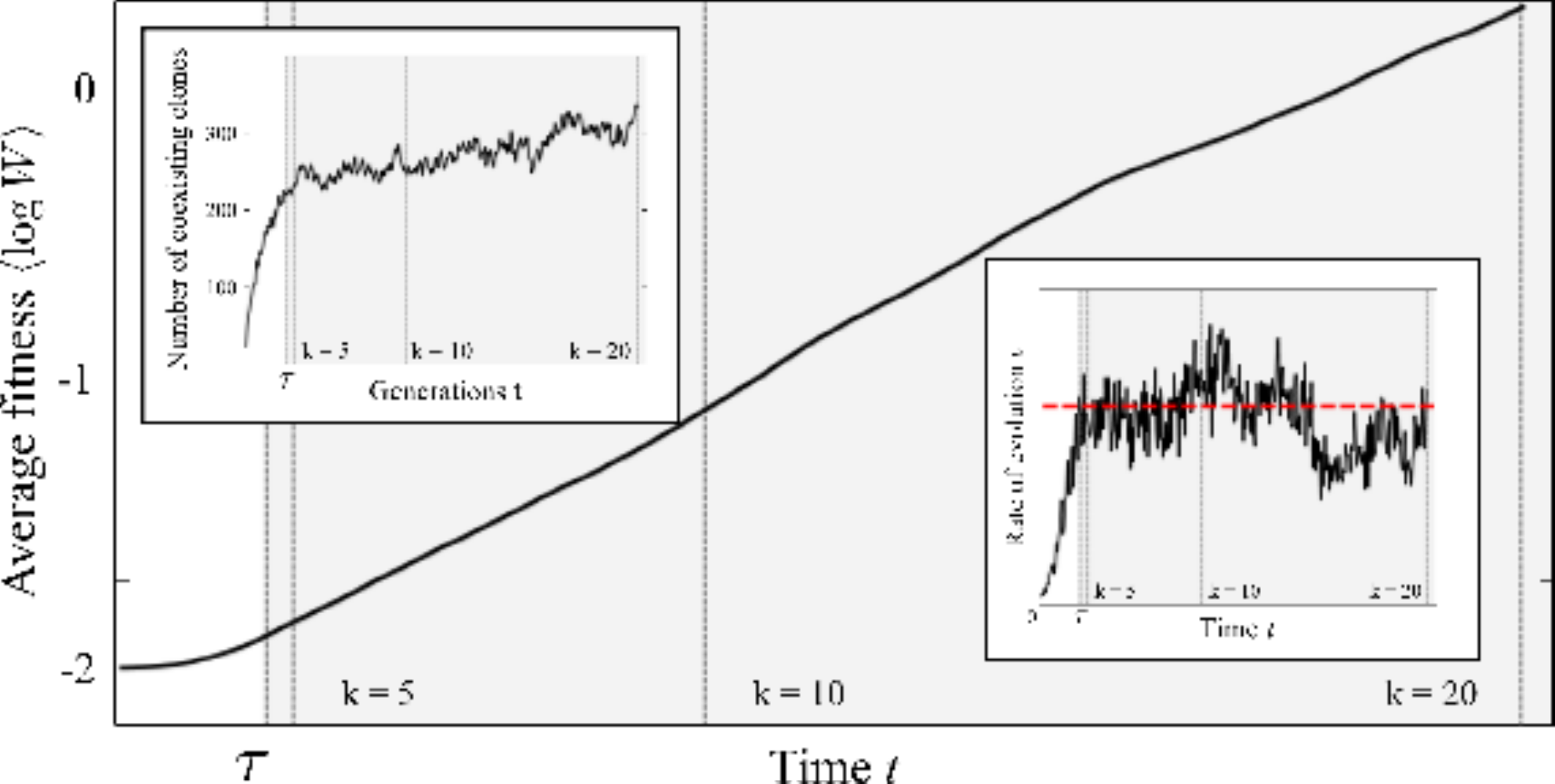}
 \caption{The logarithmic fitness $\langle \log{W(t)} \rangle$, averaged over the entire crypt population (lattice sites), increases at a constant average rate $v \cdot s_0$ after a transient time $\tau$ (see Ref.~\cite{Martens2011}). At this point, the number of coexisting clones also starts to saturate (left inset). After the saturated state, the accumulation rate carries out a random walk (right inset) around its mean value $v$ (dashed red line), given by Eq.~(\ref{eq:Vmax}), resulting in the Gaussian distribution of waiting times seen in Fig.~\ref{fig:twaithisto} (c,d). The relative width of the Gaussian distribution of waiting times scales like $\sigma / \etwait \sim (v/k)^{1/2}$, see Eq.~(\ref{eq:Gaussianwidth}). Vertical dashed lines denote the times at which $k=5,10,20$ hits have been acquired. The simulation is carried out with $s_0=0.05$, $\mu=10^{-3}$ (corresponding to a characteristic interference length $L_c\sim8$) and a neoplasm size $L=100$, until $k=20$ hits have occurred. }
\label{fig:timeevolution} 
\end{figure}

In our simulation, the neoplasm is initially devoid of mutations. There is a transient time period where the number of clones increases until the model dynamics relaxes to a state, where  newly arising and extinguishing clones stay in balance. Because of the present fluctuations, this state corresponds to a (quasi) steady state.
In Fig.~\ref{fig:timeevolution}, we illustrate the transient dynamics for clonal interference by considering the time evolution of the average logarithmic fitness $\langle \log{W(t)}\rangle$. 
The transient period may be divided into three stages: (i) until the first successful mutation appears in the neoplasm ($t < \tmut$), no change in fitness is observed; (ii) first mutations arise, the number of spreading clones begins to increase and the neoplasm is gradually populated; the fitness increases first slowly and then at an accelerating rate; (iii) once clones collide, the number of clones begins to saturate (left inset). Similarly, the rate of evolution $v$ saturates and fluctuates around a constant mean saturation value (right inset), given below by Eq.~(\ref{eq:Vmax}).

The observation that the number of coexisting clones first undergoes strong growth, and then begins to saturate, gives us a hint as of how large the transient time $\tau$ might be: it appears to be related to the time until several clonal waves begin to interfere (thus, the rate of fitness increase also saturates), suggesting a time scale of order
\begin{eqnarray}\label{eq:relaxtime}
  \tau&\sim& L_c / c_0 = \left(\frac{1}{2 s_0\, \mu \,c_0^2}\right)^{1/3}.\ 
\end{eqnarray}
Additional simulations, where we have varied both the mean selective advantage $s_0$ and mutation rate $\mu$ over two decades, have indeed consistently exhibited transient times of this order.

We have observed above that the distribution of waiting times $t_k$ for clonal interference differs from the distribution for periodic selection.
In Fig.~\ref{fig:timeevolution}, after the transient time has passed, 
the accumulation rate  randomly fluctuates around a mean value (given below in  Eq.~(\ref{eq:Vmax})). 
Waiting times $t_k$ are well-fitted by a Gaussian distribution, see Fig.~\ref{fig:twaithisto} (c) and (d) (This indicates that the fluctuations in the accumulation rate are correlated over times that are short compared to the total waiting time.).
Even after observing only a low number of hits, $k=5$, where transient behavior could potentially affect the dynamics of clonal expansions, the fit is surprisingly good.
An estimate for the width of this distribution is found further below in  Eq.~(\ref{eq:Gaussianwidth}).

\subsection{Waiting time in non-structured versus spatially-structured populations}
\begin{figure}[htp!]
\centering
    \includegraphics[width=0.9\textwidth]{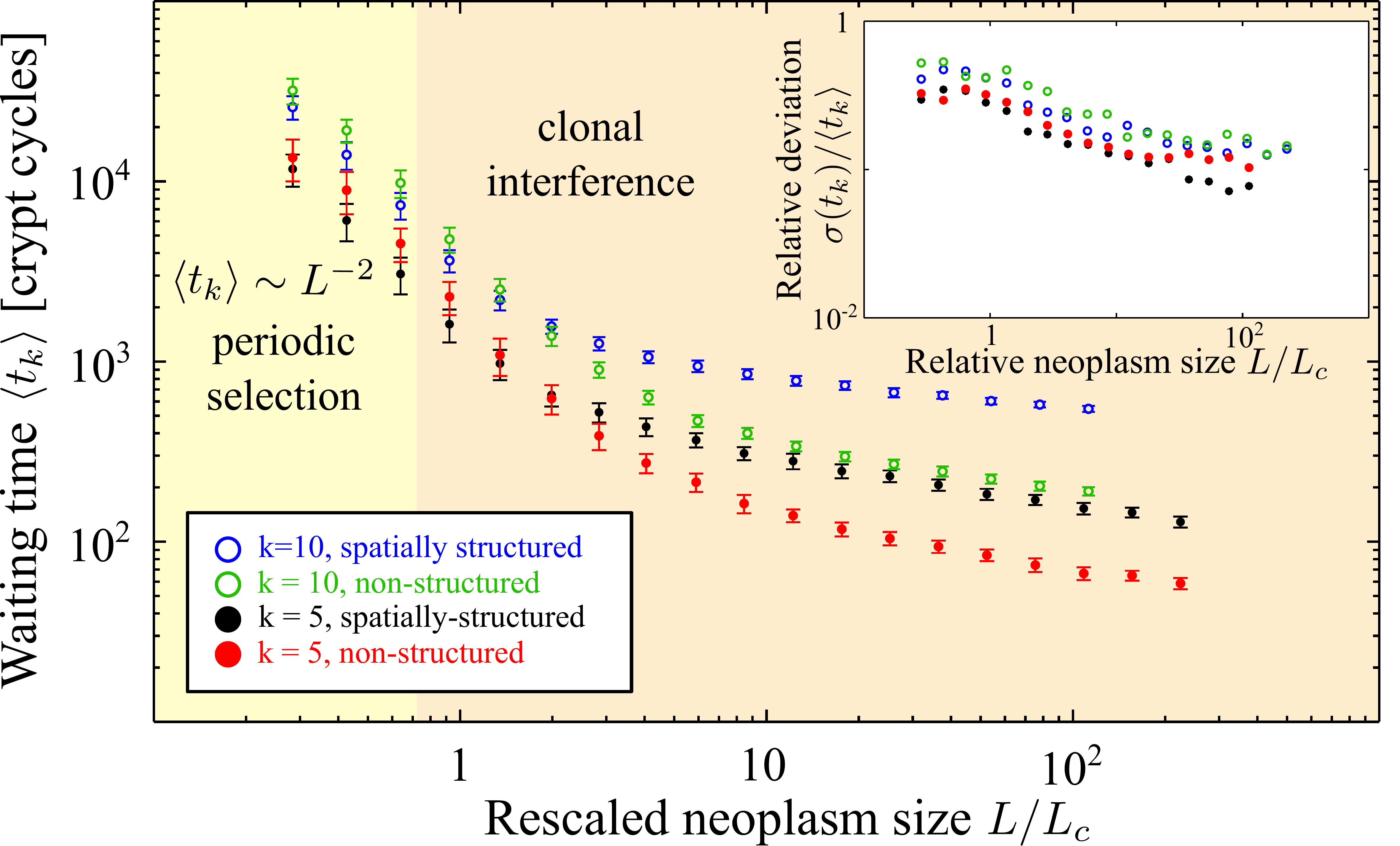}
\caption{
The average waiting time for the accumulation of $k=5$ and $k=10$ driver mutations is associated with the waiting time for cancer.
For periodic selection ($L<L_c$), the rate-limiting step is for both well-mixed and spatially extended populations given by the waiting time for the next successful mutation $\tmut$. The waiting time is thus $\etwait=k \cdot\tmut$ (the mismatch in the data for non-structured and structured populations is due to the finite size effects for small neoplasm sizes).
For clonal interference, the waiting time for $k$ hits is markedly longer for spatially structured (black) than non-structured populations (red). The discrepancy between non-structured and spatially structured populations increases for larger number of driver mutations $k$. Error bars indicate the sample standard deviations $\sigma(t_k)$. The inset displays how the relative (sample) standard deviation $\sigma(t_k)/\langle t_k\rangle$ depends on the neoplasm size $L$. 
Notice that fluctuations around the mean are smaller in the clonal interference than in the periodic selection regime (see inset).
Simulations were carried out with absorbing boundaries and exponentially distributed selective advantages. We have chosen a selective advantage of $s_0=0.05$ and a mutation rate of $\mu=10^{-4}$, such that the characteristic interference length $L_c\sim 17$ falls into the range of values we have estimated for pre-malignant tissues in the colon (see Table~\ref{tab:param}). }
\label{fig:twait} 
\end{figure}

Fig.~\ref{fig:twait} illustrates how the expected waiting time $\etwait$ depends on the rescaled (relative) neoplasm size $L/L_c$. For periodic selection, when $L<L_c$, the rate-limiting step is given by the waiting time for the next successful mutation. This time  $\tmut=1/(2s_0\mu L^2)$ is valid  for both spatially structured and non-structured populations.
Accordingly, the expected waiting time for periodic selection scales like $\etwait \sim  k\, \tmut \sim L^{-2}$ for both structured and non-structured populations. 

For large neoplasms clonal interference becomes relevant, and we observe in Fig.~\ref{fig:twait} that the waiting time for $k$ hits is shorter for non-structured crypt populations (red and green) than for spatially structured populations (black and blue). This means that, in the same time period, fewer hits can be acquired in a spatially structured  compared to a non-structured population.

This observation can be explained by the results from a study on the adaptation speed for spatially structured populations~\cite{Martens2011}. Due to the wave-like spread of clones, the accumulation rate of mutations is much lower in spatially structured when compared to non-structured populations.  
For clonal interference, the accumulation rate is given by~\cite{Martens2011}
\begin{eqnarray}\label{eq:Vmax}
 v_{} &\sim & 2s_0 L_c^2 \mu \sim (2 \,s_0^2\, c_0^2 \,\mu)^{1/3}.\
\end{eqnarray}
This rate differs in two fundamental ways from the rate for non-structured (i.e.~well-mixed) populations~\cite{Desai2007a,Beerenwinkel2007}: 
(i) while for periodic selection, the accumulation rate of mutations grows linearly with neoplasm size $v = \tmut^{-1}\sim L^2$ , it becomes strictly independent of the neoplasm sizes, constituting a strict speed limit of evolution for large neoplasms~\cite{Martens2011}; (ii) the accumulation rate  obeys a power law in the mutation rate $\mu$. 
This strongly contrasts with the speed in a well-mixed population~\cite{Desai2007a,Beerenwinkel2007}, where the accumulation rate  depends in a logarithmic way on both mutation rate and mean selective advantage. (The waiting time for the spatial case in Fig.~\ref{fig:twait} exhibits a weak dependence on the neoplasm size $L$, presumably due to the transient behavior discussed above.)
When transients do not matter or are negligible (for instance for large numbers of hits $k$), the waiting time  for $k$ hits for clonal interference can be approximated by
\begin{eqnarray}\label{eq:tkCI}
 \etwait &\approx& \frac{k}{v_{}}.\
\end{eqnarray}
A more general formula may possibly be cast into the form $\etwait \sim \tau + (k-\tilde{k}(\tau) )/v_{}$, where $\tilde{k}$ is the average number of hits acquired during the transient time $\tau$. At this time, we have no prediction for this more general case including transient behavior; we hope to address this question in a future study.


Finally, using Eq.~(\ref{eq:tkCI}), we estimate the width of the distribution of waiting times for clonal interference (Fig.~\ref{fig:PSCI} (c,d)). 
As a consequence of the random (short-time) fluctuations of the accumulation rate, the variance in waiting times grows linearly with the mean waiting time, i.e.~$\sigma^2 \propto \etwait \propto  k /v$. Thus the \emph{relative} width of the distribution scales like  
\begin{eqnarray}\label{eq:Gaussianwidth}
\frac{\sigma} {\etwait} &\propto& \sqrt{v/k}.\ 
\end{eqnarray}




\subsection{Distribution of clonal patch sizes}
\begin{figure}[htp!]
\centering
\includegraphics[width=0.9\textwidth]{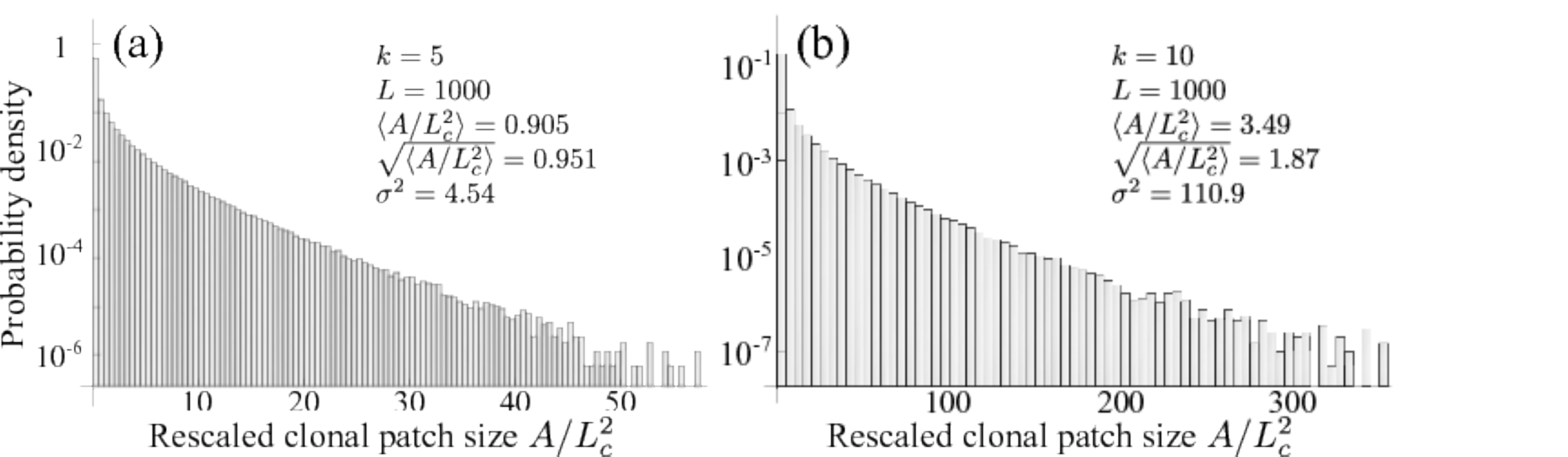}
\caption{Distribution of rescaled clonal patch sizes after $k=5$ and $k=10$ hits have occurred (clonal interference). The patch size $A$ is defined as the number of crypts carrying the same mutation. 
The mean clonal patch size is of order $L_c^2$.
Data represents 1000 realizations with a neoplasm width of $L=1000$, exponentially distributed selective advantages with mean $s_0=0.05$, and a mutation rate of $\mu=10^{-4}$ (yielding an interference length of $L_c\sim 17$, in agreement with estimates in Table~\ref{tab:param}). Clonal patches of 1 are excluded, since the patch at the end of the simulation always has size 1, thus generating an unrealistic bias.}
\label{fig:cphisto} 
\end{figure}

We investigate the size of clonal patches, defined as the number of crypts $A$ carrying the same mutation. For periodic selection, patch sizes are uniformly distributed within $[0,L^2]$, provided that sampling occurs at arbitrary times. When clones interfere, however, patch sizes are non-uniformly distributed as shown in Fig.~\ref{fig:cphisto}. 
We find that the mean clonal patch size is $\langle A\rangle\sim L_c^2$, 
justifying our above interpretation of $L_c$ as the typical size of clones.

\subsection{Genetic distance versus spatial distance}
\begin{figure}[htp!]
\centering
\includegraphics[width=0.9\textwidth]{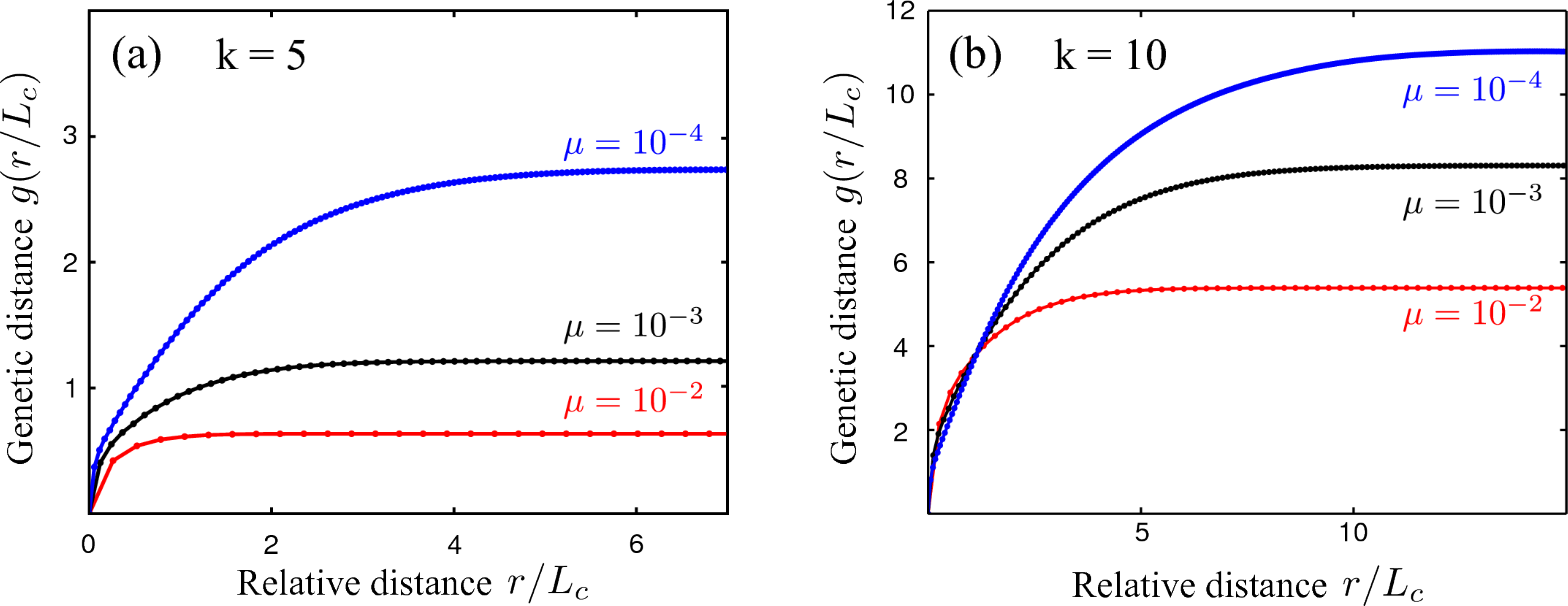}
\caption{
Genetic (Hamming) distance (number of differing mutations in two crypts) increases as a function of spatial distance over the characteristic interference scale $L_c$ (for clonal interference). The genetic distance $g(r)$ in the figure represents the average genetic distances found within a radial interval $[r,r+\Delta r[$ with binning width $\Delta r=1$, and is calculated after $k=5$ (a) or $k=10$ (b) hits have occurred. 
Each curve represents the ensemble average over 100 realizations; standard errors are too small to be seen on this plot.
Simulations were carried out for neoplasms with side length $L=316$, exponentially distributed selective fitness effects with mean $s_0=0.05$, mutation rates $\mu=10^{-4},10^{-3},10^{-2}$ per crypt and a crypt cycle, corresponding to $L_c\sim 4,8,18$.}
\label{fig:correl} 
\end{figure}

Genetic sequencing of several clones within a neoplastic tissue~\cite{Salk2009,Greaves2006} allows to measure the genetic distance between clones. Here, we investigate whether this genetic distance depends on spatial distance in the tissue.
We define the genetic distance $n$ between two crypts as the number of mutations by which they differ (Hamming distance).
Since no crypt possesses more than $k$ mutations at the end of a simulation, the genetic distance is bounded by $0\leq n\leq 2k$. 
The average genetic distance between two crypts, given a distance $(p,q)$, is then computed via
\begin{eqnarray}
 \tilde{g}_{pq} &=& \frac{1}{\Lambda_{pq}}\sum_{(i,j)} n[(i,j),(i+p,j+q)],\
\end{eqnarray}
where the sum is carried out over all lattice points $(i,j)$ such that $(i+p,j+q)$ lies within the neoplasm (i.e.~a member of the simulated lattice); $\Lambda_{pq}$ represents the number of points over which the sum is carried out for each tuple $(p,q)$. To obtain a radial correlation function $g(r)$, distances are binned into groups of $[r,r+\Delta r[$ with $r=\sqrt{p^2 +q^2}$ and $\Delta r =1$. For $(p,q)=(0,0)$, crypts are identical and we have $n[(i,j),(i,j))] = 0$ and $g(0)=0$. Since the genetic distance can at most be $n=2k$, so is also $g(r)$ bounded by $2k$.
Fig.~\ref{fig:correl} shows that the genetic distance $g(r)$ attains maximal value;  remarkably, this maximal genetic distance increases with smaller mutation rates $\mu$. The genetic distance begins to increase (i.e.~ the genetic correlation decays) on the scale of the characteristic interference length $L_c$.

\subsection{Distribution of selective advantages of sweeping mutations}
Our model assumes an exponential distribution of fitness effects for the selectively advantageous mutations. One may wonder which fitness effect distribution will be found in those clones that eventually drive tumor progression. In general, in the clonal interference regime, this distribution should deviate from the exponential distribution, because mutations with small effect will not help in the competition with other clones.

\begin{figure}[htp!]
\centering
\includegraphics[width=\textwidth]{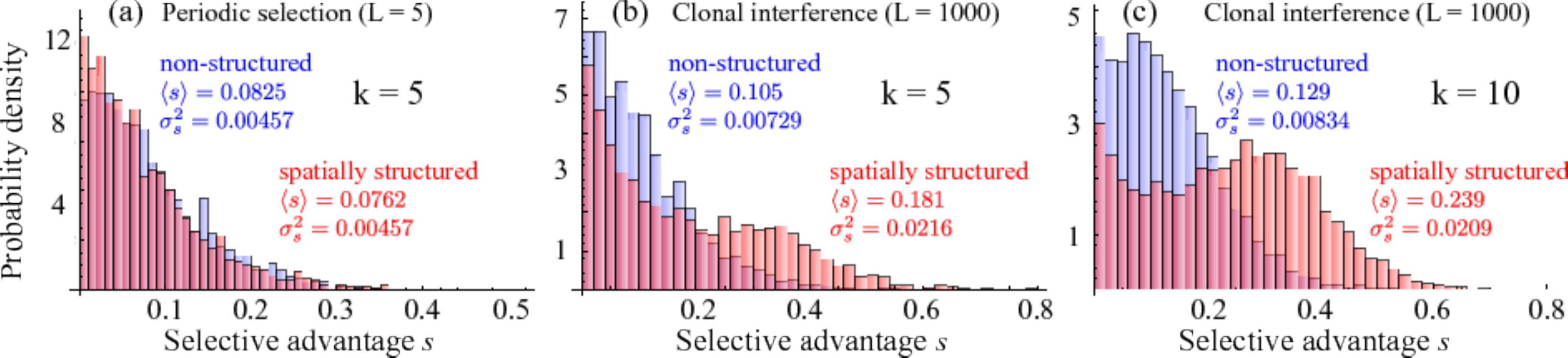}
\caption{
Distribution of fitness effects of mutations in the terminating (''winning``) clone with $k=5$ and $k=10$ hits. For periodic selection ($L=5$), the shape of the fitness distribution is similar to the exponential distribution from which effects are drawn during the simulation (a); no significant difference is discernible in the fitness distribution for non-structured (blue) and structured populations (red).
However, when clones interfere ($L=1000$), the following observations are made (panels (b) and (c)): (i) the mean $\langle s\rangle$ is larger, demonstrating that (ii) almost neutral mutations are less likely to survive than for periodic selection, and (iii) distributions for non-structured and spatially structured populations have markedly different shapes:
the distribution of fitness effects exhibits a distinct bimodal structure for the spatially structured case, and the associated mean selective advantage is approximately twice as large as in the non-structured population. 
500 Simulations have been carried out with selective fitness effects drawn from an exponential distribution with mean $s_0=0.05$ and a mutation rate of $\mu=10^{-4}$ (i.e.~$L_c\sim17$, in agreement with our estimates for colonic tissue in Table~\ref{tab:param}).
}
\label{fig:scoeffdist} 
\end{figure}

Fig.~\ref{fig:scoeffdist} shows the distribution of selective fitness effects of the mutations that have hit the ''winning`` clone.
When the simulation terminates with the appearance of $k$ hits, selective fitness advantages are sampled from the crypt where the $k$-th hit has occurred (we exclude the fitness coefficient $s$ that is due to the  terminating hit). 

For periodic selection, the distribution has a similar shape as the exponential distribution from which selective advantages are drawn during the simulation, see Fig.~\ref{fig:scoeffdist} (a). However, due to genetic drift (number fluctuations), mutations only survive with a probability of $2s$, which biases the original fitness distribution. The mean selective fitness is larger than the mean of the exponential distribution $s_0$ from which selective coefficients are drawn. 

A rather different picture is seen for clonal interference: the distribution of fitness effects has strongly changed, both in non-structured and spatially structured populations of crypts, see Fig.~\ref{fig:scoeffdist} (b) and (c). It is seen that mutations with small fitness advantage are less likely to survive for clonal interference than for periodic selection. 
As a consequence, the mean value of the fitness effects is larger for clonal interference than for periodic selection.
The loss of nearly neutral mutations is often discussed in the context of (nearly) neutral passenger mutations~\cite{Bozic2010}. 

Interestingly, the distributions for the spatially structured and the well-mixed case are markedly different. The distribution for the spatially structured case exhibits a second local maximum in the distribution for large values of $s$. This hump is even more pronounced for longer simulation times, i.e.~when $k=10$ hits have occurred (Fig.~\ref{fig:scoeffdist}(c)). 
In the spatially structured population, the mean value $\langle s\rangle$ is almost twice as large. 
The influence of clonal interference on the survival of selectively advantageous mutations has recently been studied theoretically in detail for non-structured populations~\cite{Fogle2008};
remarkably, we find here that in spatially structured populations, large effect mutations are more likely to survive than almost neutral mutations, when compared to non-structured populations.


\subsection{Estimating $\lint$ for pre-cancerous tissue}
In the following, we estimate the characteristic interference length $L_c$ for pre-malignant conditions related to colon cancer, for which parameters are available, to assess whether clonal interference is likely. 
Based on these parameters, we provide an estimate for a lower and an upper bound of the interference length $\lint$ and compare them with typical sizes of neoplasms in colon cancer.

\paragraph{Units of length and time.}
We measure length in units of crypt spacings, and time in units of crypt cycles. 
Crypts in the colon have typically an average diameter of approximately $\delta\sim 0.05$ mm~\cite{Tamura2002}. 
Crypts need not be packed perfectly dense, and lie at most a few orifice widths $\delta$ apart (Fig.~\ref{fig:cryptbifurcation} (b)). 
We therefore consider a  range for the crypt displacement $\Delta$ of about two to four crypt diameters, i.e.~$\Delta\sim 2\delta \ldots 4 \delta$. 
Typical turn-over times for crypts in diseased colonic tissue lie in the range of 50 to 200 days~\cite{Cheng1986};  
this rate is significantly higher than in normal healthy tissue where crypts have been observed to divide only once every 9 to 17 years~\cite{Cheng1986,Totafurno1987}. 

\paragraph{Mutation rate.}
In our model, we consider the crypts as local proliferative units, and
account for the kinetics of crypts rather than individual cells. In what follows, we therefore estimate the mutation rate $\mu$ per crypt and crypt cycle using a simple formula (Eq.~\ref{eq:murate}).

Estimates for mutation rates in human cells lie in the range of $u=10^{-7}\ldots 3\times10^{-6}$ per gene and cell division~\cite{Beerenwinkel2007,Bozic2010,Frank2007book,Calabrese2010,Araten2005}. However, the human genome consists of a large number of genes, but not all mutations necessarily lead to cancer. 
Different mutation patterns have been observed in different cancer types, and  are subject to considerable variation between patients~\cite{Sjoeblom2006,Greenman2007}.
Thus, there exists a subset of target genes that ultimately may lead to the growth of carcinoma, for instance by deactivation of specific tumor suppressor genes.  Analysis of genotype data obtained from colorectal and breast cancer patients~\cite{Sjoeblom2006} leads to an estimate of $d \sim 40$ to $100$ target genes. Only a fraction of these target genes are altered in a patient during progression to cancer~\cite{Beerenwinkel2007,Bozic2010}. An estimate for the effective mutation rate for target genes is given with  $u\times d$.

Crypts in the colon typically harbor around 2000 cells of various type~\cite{Calabrese2010,Nicolas2007}, but only a small number of long-lived stem cells. The turnover of cells within the crypt is high~\cite{Wright1984}, and stem cells are the only cell type that remains in the crypt long enough to acquire sufficient carcinogenic mutations. 
Colonic crypts typically harbor $\nu = 8 \ldots 20$ stem cells~\cite{Calabrese2010,Totafurno1987,Nicolas2007}.
Accordingly, we estimate the mutation rate per crypt and crypt cycle as follows,
\begin{eqnarray}\label{eq:murate}
 \mu &=& u \, d \, \nu \, \frac{T}{\tau}= 5\times 10^{-4} \ldots 0.4 \quad\textrm{per crypt and crypt cycle},\
\end{eqnarray}
where $d$ is the number of target genes, $\nu$ the average number of stem cells in a crypt, $T$ the length of a crypt cycle and $\tau$ the duration of a cell division.

\paragraph{Velocity of adaptation waves.}
Details of the crypt bifurcation process are not well understood~\cite{Merlo2006}, and measurements of the expansion velocity of clones are not available at this time. To estimate their velocity, we use the velocity relation valid for our model, i.e.~$c\sim\sqrt{\Delta W}$ where we let $\Delta W = s_0$. The front of adaptive waves thus move faster the larger the fitness difference $\Delta W$ is along two neighboring regions inhabited by different clones.

The selective advantage of mutations is generally difficult to measure~\cite{Eyre-Walker2007}, and values based on experimental data from neoplastic tissue are unavailable.  Some studies have used simulation models to infer the mean selective advantage $s_0$ by matching the expected waiting time $\langle t_k\rangle$~\cite{Beerenwinkel2007,Bozic2010}, based on a number of assumptions.
Similar to previous studies~\cite{Beerenwinkel2007,Bozic2010}, we consider a relatively large range of values of $s_0 \sim 0.001\ldots 0.1$.
Using these values, we have an adaptive wave velocity of
\begin{eqnarray}
 c & \sim & 0.03 \ldots 0.3 \textrm{ crypts per crypt cycles}.\
\end{eqnarray}


Due to the cube root law in Eq.~(\ref{eq:Lc}), the interference length $L_c$ depends very weakly on the key parameters (wave speed $c_0$, mutation rate $\mu$  and mean fitness effect $s_0$). The rather large value range for $s_0$ considered here does therefore not strongly affect the estimate for $L_c$. In particular, with a clonal wave speed with $c\sim\sqrt{s}$, we have $L_c\sim s_0^{-1/6}$, rendering estimates even more insensitive to variations in $s_0$. For instance, there is evidence that specific cancer related genes may yield a very large fitness coefficient of $s\sim 1$~\cite{Pritchard1998}; however, even such a large selective advantage does not strongly alter our estimate for $L_c$, and only decreases the lower estimated boundary to $L_c\sim 1$.

\paragraph{Interference length $\lint$ and size of neoplasms.} 
Using Eq.~(\ref{eq:Lc}) and the parameter values from Table~\ref{tab:param}, we estimate a lower bound of the characteristic interference length with $s_0=0.1$, $\mu=0.4$, and an upper bound with parameters $s_0=10^{-3}$, $\mu=5\times 10^{-4}$. 
We thus find the following range for the characteristic interference length,
\begin{eqnarray}
 \lint &\sim & 1.5 \ldots 30 \textrm{ crypts}.\
\end{eqnarray}
This range includes mean patch sizes measured in a study on clonal expansions in the colon~\cite{Greaves2006}.
In our simulations, we have chosen parameters resulting in a characteristic interference length $L_c\sim 17$ (except when varying $L_c$ or stated otherwise).

\begin{table}
\small
\centering
 \begin{tabular}{l|c|cc|l}
  \hline \hline
  Parameter & symbol & min & max & References\\
 \hline
  
 mutation rate  [per cell division per gene] &$u$& $10^{-7}$&$3\times10^{-6}$& \cite{Calabrese2010,Beerenwinkel2007}\\
 number of stem cells in crypts   &$\nu$  & $8$ & $20$ & \cite{Nicolas2007,Totafurno1987,Calabrese2010}\\
 number of target genes &$d$& $40$&$100$&\cite{Bozic2010,Beerenwinkel2007}\\
 crypt bifurcation time [days]&$T$&&&\\
  \hspace{.1cm}ulcerative colitis:&  & 50 & 105 &\cite{Cheng1986}\\
  \hspace{.1cm}Crohn's disease: & &100 & 200 &\cite{Cheng1986}\\
  \hspace{.1cm}multiple polyposis: &  &110 & 230&\cite{Cheng1986}\\
 stem cell division time [days]&$\tau$&3 &3&\cite{Totafurno1987}\\
 mutation rate [per crypt and cycle]& $\mu$ & $5\times 10^{-4}$ & 0.4 &[see text]\\ 
\hline
  mean selective advantage & $s_0$ &$10^{-3}$&$0.1$&\cite{Beerenwinkel2007,Bozic2010}\\
 wave velocity [crypts / cycle] &$c_0$&0.032&0.32&[see text]\\
 \hline
 
 interference length [crypts]& $\lint$ & 1.5& 30&[see text]\\
\hline
 neoplasm diameter [crypt distance] & $L$& 50 & 500& [see text]\\
 \hline \hline \hline
 \end{tabular}
 \caption{Adaptive clonal waves collide at the characteristic interference length scale $\lint$. Here we list parameters as obtained for the human diseased colon which were used to estimate the interference length $\lint$. When the size of the neoplasm size $L$ exceeds the characteristic interference length $\lint$, adaptive waves collide and compete with each another in going to fixation.   }
 \label{tab:param}
\end{table}

To assess whether clonal interference is likely to occur, we compare our estimate of $L_c$ with the typical neoplasm size $L$ in the colon.
Crypts are relatively densely packed and have an orifice width of around $\delta=0.05$ mm~\cite{Tamura2002}. Assuming that the crypt spacing is of the order $\Delta\sim 2\delta \ldots 4 \delta$, we get a crypt density of 5 to 10 crypts per mm (this is also in line with a recent measurement of 16 crypts per square mm in the sigmoid colon~\cite{Cheng1984}). 
The diameter of an adenoma lies in the range of 10 to 50 mm~\cite{Nusko1997}, corresponding to a diameter ranging from 50 to 500 crypts.
These neoplasm diameters are larger than the characteristic interference length we have estimated above. We therefore conclude that clonal interference likely occurs in neoplasms related to colon cancer, and possibly in other pre-cancerous tissues with spatial character. 

Parameters for Barrett's segment are subject of current research; one may however speculate that parameters  are similar to those in the colon, because the epithelial structure have similar (intestinal epithelial) character. Barrett's segment extends typically over a length of 1 cm to 10~cm~\cite{Rudolph2000}, and has a circumference of around 8 cm (quarter-dollar coin). Assuming a crypt density similar to the diseased colon, one finds an average width of 150 to 900 crypts and thus Barrett's segment may also be subject to clonal interference.


\section{Discussion}
The modeling of carcinogenesis has a long history since its introduction more than 50 years ago~\cite{Armitage1954,Armitage1957}. 
Recently, several studies have focused on the waiting time to cancer~\cite{Beerenwinkel2007,Bozic2010,Foo2011}, which may be defined as the time until a critical number of hits (driver mutations) are accumulated and initiate the growth of carcinoma.
While most of these theoretical studies have considered well-mixed populations exclusively, our study considers neoplastic progression in tissues with pronounced spatial structure.

Although it is reasonable to assume that somatic cell populations are well-mixed in certain cancers, such as blood cancer, 
other precancerous tissues have distinct spatial structure, which has important implications for the way in which clones expand. 
For instance, neoplasms in colon cancer and in Barrett's esophagus are organized into proliferative units called crypts~\cite{Merlo2006,Potten1998,Totafurno1987}, which divide the epithelial tissue into sub-populations of stem cells. In the colon, evidence has recently been found that clones expand via crypt bifurcation~\cite{Salk2009,Greaves2006}. 

Clonal expansions in spatial structures can be modeled by the spread of adaptive waves, which were first introduced by Fisher~\cite{Fisher1937}. Such waves travel at a constant speed through the neoplasm~\cite{Fisher1937,Kolmogorov1937,Hallatschek2009}. This ''ballistic`` spreading is markedly slower than in non-structured populations, where clones obey (exponential) Malthusian growth. As a consequence of this slower progression, clonal waves are more prone to interfere, and may lead to slower adaptation speeds in comparison to non-structured (well-mixed) populations~\cite{Martens2011}. 
We found that spatial structure - compared to non-structured cell populations assumed in other studies~\cite{Beerenwinkel2007,Bozic2010} - increases the waiting time to acquire a given number of $k$ hits (i.e.~driver mutations). The discrepancy of waiting times between well-mixed and spatially structured cell populations grows for larger hit numbers.
A frequently debated question is how many hits (driver mutations) are required to initiate cancer, and figures range from 2, 5 and up to 20 hits~\cite{Beerenwinkel2007,Wood2007,Sjoeblom2006,Jones2008a,Parsons2008}.
Our finding shows that the number of hits that can be acquired during a human lifetime must be lower for spatially structured neoplasms than when assuming a well-mixed structure of the neoplasm.
Parameter estimates, such as for the mean selective fitness advantage found by matching the hit number $k$, may thus need to be revised for spatially structured cancers with spatial structure, too. 

Our spatial model allows us to predict characteristic structural features of neoplastic tissue. When clones interfere, our model predicts that neoplasms are subdivided into clonal patches, which are non-uniformly distributed with a typical average size $L_c^2$ (Eq.~(\ref{eq:Lc})). This is consistent with a recent experimental study which finds that mutated crypts form clusters in the human colon~\cite{Greaves2006}.
Moreover, our investigations on the genetic (Hamming) distance between crypts as a function of spatial distance have revealed that the genetic distance starts to increase over the characteristic interference length $L_c$. 
Remarkably, the maximal genetic distance is found to increase with decreasing mutation rate.
Using genetic sequencing methods, it should be possible to measure genetic distances between samples from different locations in neoplastic tissue, and to test for the features of clonal interference we report here. 
Indeed, recent studies on clonal expansions in the human colon~\cite{Salk2009,Greaves2006} have introduced promising methods to measure the size of clonal patches using sequencing methods.

The distribution of waiting times is different for clonal interference than for periodic selection, where clonal waves sweep strictly sequentially. While the waiting times obey a Poisson-like distribution for periodic selection, they follow a Gaussian distribution for clonal interference.
The waiting time for cancer progression is (experimentally) often unknown, since the precise time of initiation of a pre-malignant condition stays undetected. It may be possible, though, to tie the distribution of waiting times to the age distribution of cancer outbreak~\cite{HerreroJimenez2000}, using statistical models.

When clones interfere, mutations with small effect are less likely to survive~\cite{Fogle2008} than for periodic selection, because they contribute little to the competition between clones. Remarkably, for spatially structured populations, we have found that this effect is even stronger, and that the distribution of fitness effects of sweeping mutations has a bimodal shape. It will be interesting to investigate this phenomenon in a future study, in the context of passenger mutations and their role in cancer progression~\cite{Bozic2010}.

Two different paradigms of asexual evolution - periodic selection and clonal interference - have been suggested as potential models for cancer progression~\cite{Graham2010}. Ultimately, knowing the appropriate model of cancer progression not only improves our theoretical understanding, but will also be very valuable for the interpretation of genotyping data obtained from biopsies. We have discussed multiple characteristic features of clonal interference and differences to periodic selection that may assist in deciding this question. 

We have estimated the characteristic interference length $L_c$ for neoplasms in the colon. 
When $L_c$ -- the the typical length scale at which adaptive waves interfere~\cite{Martens2011} -- is smaller than the neoplasm size $L$, clones interfere. Using available parameters for the diseased colon, we have estimated the characteristic interference length to be in the range of 1 to 30 crypts. This is considerably smaller than neoplasms in the colon, and we therefore expect that clonal interference likely occurs, and that it also may play an important role in other cancers with spatial structure.
Our estimates depend on rough estimates of the selective advantages of driver mutations, as well as the spreading velocity of clones by crypt bifurcation.
However, the interference length depends only weakly on parameters (with a cube root law), and therefore uncertainties regarding parameters only have a small influence on our estimate of $L_c$. Due to this insensitivity of $L_c$ on parameters, it should be possible to assess $L_c$ for other cancers with spatial character, too. 

It is interesting to think about the role of metastasis in the context of clonal interference. When metastasis occurs at a late stage of cancer development, cancer cells disperse to far distant organs of the body and proliferate there. This means that these cells are capable of spreading in quite different environments and that their proliferation relies on a variety of mechanisms. However, if one considers only a single organ (e.g.~colon), or 'mid-range' migration of cells within the same or neighboring neoplasms, metastasis takes place within identical environments. 
Long-range migration (i.e.~any form of spatial mixing) strongly increases the speed of evolution~\cite{Martens2011}: by generating multiple seeds far off the originating clone, the effective cumulated sweeping speed becomes larger. As a consequence, clones are less likely to interfere, and the accumulation rate of mutations increases. One may speculate that this is the reason why the phenotype of metastasis confers a selective advantage.

Mortality rates in many cancers have fallen only slowly since the declaration of the ‘war on cancer’ in 1971~\cite{Sporn1996}; novel approaches to both therapy and prevention are needed. 
Currently, genotyping data from biopsies becomes increasingly available. Better spatial and temporal resolution in genetic data may significantly improve our understanding of clonal expansions and prediction of cancer progression. 
By demonstrating the important role of spatial structure in progression of some cancers, we hope to contribute an important element to the ongoing research and fight on cancer.

\appendix
\section{Waiting time distribution for $k$ hits for periodic selection}\label{appA}
For periodic selection, the accumulation of $k$ hits may be understood in terms of $k$ Poisson processes.
The probability distribution for the waiting time for $k$ hits is given by
\begin{eqnarray}
 p_k(t) &=&-\frac{\partial}{\partial t} P_{<k}.\
\end{eqnarray}
The probability of at most $k-1$ hits having occurred is given by 
\begin{eqnarray}
 P_{<k}(t) &=& \sum_{j=0}^{k-1}\pi(j),\
\end{eqnarray}
 where $\pi(j)=t^j e^{-t}/j!$ is the Poisson distribution. The resulting distribution,
\begin{eqnarray}
 p_k(t) &=&\frac{t^{k-1}e^{-t}}{(k-1)!}=\pi(k-1),\
\end{eqnarray}
matches the simulated rescaled waiting times $t_k/\tmut$ very  well, see Fig.~\ref{fig:twaithisto} (a,b).

\ack 
We would like to thank Paul Sniegowski for helpful discussions and Hedvika Toncrova  for useful comments on the manuscript, and Amir Bar for
pointing out an error in our first version of the waiting time distribution in the periodic selection regime.
This work was supported in part (R.K. and C.C.M.) by Research Scholar Grant \#117209-RSG-09-163-01-CNE from the American Cancer Society and NIH grants P01 CA91955, P30 CA010815, R01  CA149566 and R01 CA140657.

\newpage
\section*{References}

\end{document}